\documentclass[12pt]{article}
\begin{document}

\newcommand{\vv}{\vskip5pt\noindent}
\newcommand{\p}{\par\noindent}

\newcommand{\RE}{\mathsf R}
\newcommand{\C}{{\mathsf C}}

\newcommand{\phireg}{\phi_{reg}}
\newcommand{\varphireg}{\varphi_{reg}}
\newcommand{\dphireg}{\dot \phi_{reg}}
\newcommand{\vphireg}{\varphi_{reg}}

\newcommand{\E}{{\cal E}}
\newcommand{\CH}{{\cal H}}
\newcommand{\Q}{{\cal Q}}
\newcommand{\J}{{\cal J}}
\newcommand{\CS}{{\cal S}}
\newcommand{\F}{{\cal F}}

\newcommand{\LD}{L^2(\RE^3)}
\newcommand{\HU}{H^{1}(\RE^3)}
\newcommand{\HD}{H^{2}(\RE^3)}
\newcommand{\HUP}{\bar H^{1}(\RE^3)}
\newcommand{\HDP}{\bar H^2(\RE^3)}
\newcommand{\HDPZ}{\bar H^2_0(\RE^3)}
\newcommand{\HMP}{\bar H^{1/2}(\RE^3)}
\newcommand{\HTP}{\bar H^{3/2}(\RE^3)\cap \bar H^1(\RE^3)}
\newcommand{\DUP}{\bar D^1(\RE^3)}
\newcommand{\DDP}{\bar D^2_{\alpha}(\RE^3)}
\newcommand{\DTP}{\bar D^3_{\alpha}(\RE^3)}
\newcommand{\DDPZ}{\bar D^2_0(\RE^3)}

\newcommand{\nr}{{nr}}
\newcommand{\WNR}{ W_\alpha^{nr}}
\newcommand{\DDNR}{[D^2_\alpha(\RE^3)]_{nr}}
\newcommand{\DUNR}{[D^1(\RE^3)]_{nr}}
\newcommand{\LDNR}{[\LD]_{nr}}
\newcommand{\DUPNR}{[\bar D^1(\RE^3)]_\nr}
\newcommand{\DDPNR}{[\bar D^2_{\alpha}(\RE^3)]_\nr}
\newcommand{\BWNR}{\bar W_\alpha^{nr}}
\newcommand{\Deltanr}{\Delta_{\alpha}^{nr}}

\newcommand{\qed}{\hfill\vbox{\hrule\hbox{\vrule\vbox to 7 pt {\vfill\hbox to
         7 pt {\hfill\hfill}\vfill}\vrule}\hrule}\par}
\newcommand{\uno}{{\mathsf 1}}


\vskip 80pt
\centerline{\bf{\sf WAVE EQUATIONS WITH POINT INTERACTIONS}}
\centerline{\bf{\sf IN FINITE ENERGY SPACES}}
\vskip 40pt
\centerline{Massimo Bertini}\smallskip
\centerline{\it Dipartimento di Matematica, Universit\`a di Milano}
\centerline{\it Via Saldini 50, I-20133 Milano, Italy} 
\smallskip
\centerline{E-mail: {\tt bertini@mat.unimi.it}}
\vskip 20pt
\centerline{Diego Noja}\smallskip
\centerline{\it Dipartimento di Matematica, Universit\`a di Milano}
\centerline{\it Via Saldini 50, I-20133 Milano, Italy} 
\smallskip
\centerline{E-mail: {\tt noja@mat.unimi.it}}
\vskip 20pt
\centerline{Andrea Posilicano}
\smallskip
\centerline{\it Dipartimento di Scienze, Universit\`a dell'Insubria}
\centerline{\it Via Valleggio 11, I-22100 Como, Italy}\smallskip
\centerline{E-mail: {\tt posilicano@mat.unimi.it}}

\vskip 150pt
Given the abstract wave equation 
$\ddot\phi-\Delta_\alpha\phi=0$,
where $\Delta_\alpha$ is the Laplace operator with
a point interaction of strength $\alpha$, we define and study $\bar W_\alpha$, the associated wave generator in the phase space
of finite energy states. We prove the existence of the phase flow
generated by $\bar W_\alpha$, and describe its most relevant
properties with particular emphasis on the associated symplectic
structure and scattering theory.

\vfill\eject\p
{\bf{\sf I. INTRODUCTION.}}\vv
To introduce the problem we begin with a well known example.
Given the free scalar, {\it zero mass}, wave equation
\begin{equation}
\ddot \phi-\Delta \phi=0\, ,
\end{equation}
the usual attitude in the literature is to search the solutions
in the real Sobolev-Hilbert space $H^2(\RE^3)$; in order to fix the
notations we recall that $H^s(\RE^3)$, $s\in\RE$, is defined as the set
of tempered distributions with a Fourier transform which is square
integrable w.r.t. the measure with density $(1+|k|^2)^{s}$. 
This is a standard mathematical
choice but not the more natural one. In fact,
equation (1) can be written in the first order form
\begin{equation}
\dot \psi= W \psi\, ,
\end{equation}
where the linear operator
$$
W:H^2(\RE^3)\oplus H^1(\RE^3)\rightarrow H^1(\RE^3)\oplus  L^2(\RE^3)
$$
is defined as
\begin{equation}
W
\left(
\begin{array}{c}
\phi
\\
\dot \phi
\end{array}
\right)=
\left(
\begin{array}{cc}
0 & \uno
\\
\Delta & 0
\end{array}
\right)
\left(
\begin{array}{c}
\phi
\\
\dot \phi
\end{array}
\right)\ .
\end{equation}
Here $\Delta$ is the usual Laplace operator viewed as a self-adjoint
operator on $\LD$.
It
is well known that equation (2) generates a strongly continuous one
parameter group of evolution $$U^t:H^1(\RE^3)\oplus
L^2(\RE^3)
\rightarrow H^1(\RE^3)\oplus  L^2(\RE^3)\, . $$
This group
group is energy preserving, i.e. there exists an energy form $$
{\cal E}(\phi,\dot \phi)=\frac{1}{2}\left(\|\dot \phi\|_2^2+\|\sqrt{-\Delta}
\phi\|_2^2\right)\, , $$ coinciding with the Hamiltonian of the
system, preserved by the flow. Moreover $U^t$ constitutes a
group of canonical transformations w.r.t. the symplectic form
$$
\omega\left((\phi,\dot \phi),(\varphi,\dot \varphi)\right):=
\langle\phi,\dot \varphi\rangle-\langle\varphi,\dot \phi\rangle
$$
($\langle\cdot,\cdot\rangle$ denoting the usual scalar product on
$\LD$) and $W$ is nothing but the Hamiltonian vector field
corresponding, via $\omega$, to $\cal E$.\par As the form of
Hamiltonian $\E$ suggests, a more natural domain for the study of the
system described by (3) is the space of the finite energy states,
which is larger than the original one, because the first component
$\phi$ of such a state is not necessarily square integrable, as
instead is implicit in the standard Sobolev environment recalled
above.  This more suitable description goes as follows.\par Let us
define (general and more complete definitions will be given in the
following section) $\HUP$ as the completion of the space
$C^{\infty}_0(\RE^3)$ in the norm $\|\sqrt{-\Delta}\,\phi\|_2$. 
Now it is possible to define the new operator $\bar W$ on $\HUP\oplus
L^2(\RE^3)$, 
the Hilbert space of
finite energy states, by
\begin{equation}
\bar W:\HDP\oplus \HU\rightarrow \HUP\oplus L^2(\RE^3)\,,\qquad
\bar W(\phi, \dot \phi):=(\dot \phi, \Delta \phi)\ .
\end{equation}
where 
$$
\HDP:=\{\phi \in \HUP : \Delta \phi \in L^2(\RE^3)\}\,.
$$
It is an easy matter to verify that $\bar W$ is a skew-adjoint operator
(see e.g. [1, thm. 2.1.2], [2, \S XI.10]) so that due to Stone
theorem it defines a strongly continuous one parameter group of
evolution $$ \bar U^t:\HUP\oplus L^2(\RE^3) \rightarrow \HUP\oplus
L^2(\RE^3) $$ which is trivially energy preserving, just because the
energy coincides with the norm of the Hilbert space, and the flow is
given by a group of isometric operators. This procedure generalizes to
the case in which one considers an abstract wave equation with a
positive self-adjoint operator in the place of $-\Delta$ (see [3], [4,
\S 8]). \par Here we consider and study in detail the case in which
$-\Delta$ is replaced by $-\Delta_{\alpha}$, the Laplace operator with
a point interaction of strength $\alpha$ (see section II for its precise
definition), and construct the corresponding wave generator $\bar W_\alpha$ ; since
$-\Delta_{\alpha}$ is not positive when $\alpha<0$ one can not directly
use the results appearing in [3] and [4].\par
The abstract wave equation
corresponding to $\Delta_\alpha$, i.e.
\begin{equation}
\ddot \phi-\Delta_{\alpha} \phi=0\, ,
\end{equation}
was introduced for the first time in [5]. There, when $\phi$ is vector-valued
and when $\alpha=-\frac{mc}{e}$ ($m$ the phenomenological mass, $c$ the
velocity of light, $e$ the electric charge), it is shown that (5) describes 
the evolution of the electromagnetic field self-interacting with a
point particle in dipole
approximation (the so called linearized Pauli-Fierz model). Another
model connected with the wave equation (5), often studied in the fifties' and
sixties' literature on exact models in quantum field theory, is the so
called ``pair theory'' (see ([6]-[8] and references therein).
The classical version of this model is the regularized version of the
one we study here, and many at the time unanswered questions about its
behaviour in the ultraviolet limit find their rigorous collocation in
the present work.\par 
In [5], [9], [10] it is also shown that the Cauchy problem is well posed 
on the phase space $D^1(\RE^3)\oplus L^2(\RE^3)$, $D^1(\RE^3)\simeq
H^1(\RE^3)\oplus\RE$, (refer to section II for the definition of $D^1(\RE^3)$) 
and that the corresponding strongly continuous one
parameter group of evolution
$$
U^t_\alpha:D^1(\RE^3)\oplus L^2(\RE^3)\to D^1(\RE^3)\oplus L^2(\RE^3)
$$
preserves the energy
$$
\E_\alpha(\phi,\dot\phi):=\frac{1}{2}\,\left(\|\dot\phi\|_2^2+F_\alpha(\phi,\phi)\right)\,,
$$ 
where $F_\alpha$ denotes the bilinear form corresponding to the
self-adjoint operator $-\Delta_\alpha$. Therefore,
analogously to the case of the free wave equation, the problem of
defining (5) on the larger space of finite energy states naturally
arises. The theory of delta point interactions was
originally developped in the context of nonrelativistic quantum
mechanics (see [11] and references therein); this made natural
to use $L^2(\RE^3)$ as the underlying
Hilbert space and so, in order to
define the dynamics on the space of finite energy states, one has to
modify the original definition of $-\Delta_\alpha$, to allow the elements of 
its domain being not square
integrable.  This is done in section III where we also show
(thm. 3.1) that the operators $\bar W_\alpha$ here constructed 
generate an evolution group
$\bar U^t_{\alpha}$, a fact that, in the case $\alpha\le 0$, is not
immediately evident.  So, as an aside result, a conserved energy form
exists; this form however is not positive when $\alpha<0$, and
therefore it is not suitable to define the norm of the appropriate
phase space. \par 
In section IV we treat the Hamiltonian
formulation of the wave equations with delta interactions. Here we
solve the problem by giving a complex structure $\J_\alpha$ commuting 
with the operator $\bar W_\alpha$. This leads to an
equivalent Schr\"odinger-like
first-order formulation which, also in view of a future
quantization of the dynamical system under study, plays a key role. 
The complex
structure before mentioned is obtained considering separately the case $\alpha\le
0$ from the other case: in the strictly negative
case we obtain an invariant splitting of the phase space and the
complex structure in such a way that the Hamiltonian vector field
appears separately as a Schr\"odinger equation both on the stable and
unstable part of the phase space; in particular, on the unstable
subspace, which is finite dimensional, the Hamiltonian is that of an
harmonic repulsor, and the Schr\"odinger equation is the corresponding
ordinary differential equation, as expected. \par
The last topic treated (see section V)
is the scattering theory for the pair of operators $(\bar W_{\alpha},
\bar W)$. Being the Hilbert phase spaces for $\bar W_{\alpha}$ and
$\bar W$ different, resort has to be made to the two Hilbert space
scattering theory introduced by Kato in [4]. Using then Birman
invariance principle and the trace condition of Birman-Kuroda theorem,
we are able to prove the existence of the M\"oller wave operators and
their completeness (thms. 5.1 and 5.2). As a consequence of the machinery needed for the
definition of wave operators, one obtains a relation (see (12)-(14)) between the
evolution group (acting on the real Hilbert space of states with
finite energy) generated by $\bar W_{\alpha}$ and the unitary group
(acting on the complex Hilbert space $L^2_\C(\RE^3)$, the
complexification of $\LD$) generated by $\sqrt{-\Delta_\alpha}$ 
(in the case $\alpha<0$ one consider only the positive part of the
operator). This can be seen as a variation of the 
procedure applied in section III in the case one
uses the standard complex structure on $L^2_\C(\RE^3)$: indeed the two
structures are related in a simple way (see (10) and (11)). 
The relations (12)-(14) could also be used to define the group $\bar
U^t_{\alpha}$, the generator of which is easily seen to be $\bar
W_{\alpha}$, so providing an alternative proof of the existence of the
dynamics.  \p
\vskip 20pt\p {\bf{\sf II. PRELIMINARIES.}}\vv 
We start by giving
definitions and main properties of the Sobolev type spaces needed in
the sequel, and to which we made reference in the introduction.
We define the
family of pre-Hilbert spaces ${\tilde H^s(\RE^3)}$, $s\in\RE$, as the
set of tempered distributions with a Fourier transform (denoted by
$\hat{\quad}$ or by $\F$) which is square integrable w.r.t. to the
measure with density $|k|^{2s}$. The scalar product is defined as
$$
\langle\phi_1,\phi_2\rangle_s:=\int_{\RE^3}
dk\,|k|^{2s}\hat\phi_1(k)\,\hat\phi_2(k)\ .
$$
Note that, when $s>0$, $H^s(\RE^3)\subset {\tilde H^s(\RE^3)}$
and ${\tilde H^{-s}(\RE^3)}\subset H^{-s}(\RE^3)$, the
embeddings being continuous. Since $\left|k\right|^{-2s}$ is locally
integrable for any $s<3/2$,  
$$
\forall s<\frac{3}{2}\,,\qquad L^2(\RE^3,|x|^{2s}dx)\subset \CS'(\RE^3)\,,\qquad
{\tilde H^{s}(\RE^3)}\equiv \F^{-1}(L^2(\RE^3,|k|^{2s}dk))
$$
and thus ${\tilde
H^{s}(\RE^3)}$ is complete for any $s<3/2$ and coincides with the usual Riesz
potential spaces (see e.g. [12, \S 7.1.2]). 
\par
We can then define the isomorphism ($r-s<3/2$)
$$
(-\bar\Delta )^{s/2}:{\tilde H^r(\RE^3)}\to{\tilde H^{r-s}(\RE^3)}\,,\qquad
\F((-\bar\Delta )^{s/2}\phi)(k):=|k|^{s}\hat\phi(k)\ .
$$
Our notation is justified by observing that, in the case  $0<r=s<3/2$, $(-\bar\Delta)^{s/2}$ coincides with the closure of 
$(-\Delta)^{s/2}:H^s(\RE^3)\to\LD$.\par 
Since, contrarily to what happens for the usual Sobolev chain
$H^s(\RE^3)$, ${\tilde H^r(\RE^3)}$ is not included in ${\tilde H^{s}(\RE^3)}$ when 
$r>s$, we also define the sequence of spaces
$$
{\bar H^{n}(\RE^3)}:=\bigcap_{k=1}^{n}{\tilde
H^k(\RE^3)}\equiv \F^{-1}\left(\,\bigcap_{k=1}^{n} L^2(\RE^3,|x|^{2k}dx)\right)\ .$$
Obvioulsy ${\bar H^{n}(\RE^3)}$ is a Hilbert space with
norm 
$$\|\phi\|_{\bar H^n}:=\left(\,\sum_{k=1}^{n}\|(-\bar\Delta )^{k/2}\phi\|^2_2\right)^{1/2}\,.$$ 
We come now to point interactions; for their general theory of we refer to [11];
here we confine ourselves to
the essential definitions and results.  The operator
$-\Delta_{\alpha}$ describing a standard point interaction at the
origin with strength $\alpha$ is defined as follows. Let us introduce
the dense linear subspace of $L^2(\RE^3)$
\begin{eqnarray}
&&D^2_{\alpha}(\RE^3):=\nonumber\\
&&\left\{\phi\in  L^2(\RE^3)  \, :\, \phi=
\phi_{\lambda}+Q_{\phi}G_{\lambda},\ \phi_{\lambda} \in  H^2(\RE^3),
\left(\alpha +
{{\sqrt{\lambda}}\over {4\pi}}\,\right)  Q_{\phi}= \phi_{\lambda}(0)
\right\},\nonumber
\end{eqnarray}
where $0<\lambda\not=-\hbox{\rm sign($\alpha$)}\,(4\pi\alpha)^2$ and
$$
G_\lambda(x)={{e^{-\sqrt {\lambda}|x|}}\over {4\pi |x|}}\ .
$$
The Laplacian with a point interaction with strength $\alpha$ is the
operator
$$
-\Delta_{\alpha}:D^2_{\alpha}(\RE^3) \to L^2(\RE^3)\,,\qquad
-\Delta_{\alpha}\phi:=-\Delta\phi_{\lambda}-\lambda Q_\phi G_\lambda\ . $$
Its resolvent is given by
$$
(-\Delta_\alpha+\lambda)^{-1}=(-\Delta+\lambda)^{-1}+\left(\alpha+
{{\sqrt {\lambda}}\over {4\pi}}\,\right)^{-1}G_\lambda\otimes G_\lambda\,,
$$
where $G_\lambda\otimes G_\lambda(\phi):= \langle G_\lambda,
\phi\rangle\, G_\lambda$.\p
The bilinear form corresponding to
$-\Delta_\alpha$ has domain $D^1 (\RE^3)\times D^1 (\RE^3)$,
$$
D^1 (\RE^3):=\{\phi\in  L^2(\RE^3) \ :\ \phi=\phi_{\lambda}+Q_\phi G_{\lambda}\ ,
\phi_{\lambda} \in  H^1(\RE^3),\ Q_\phi\in \RE  \}\,,
$$
and is defined by
$$
F_{\alpha}(\phi, \varphi):=
\langle(-\Delta+\lambda)^{1/2}\phi_\lambda,(-\Delta+\lambda)^{1/2}\varphi_\lambda\rangle
-\lambda\langle\phi,\varphi\rangle
+\left(\alpha +{{\sqrt {\lambda}}\over {4\pi}}\,\right)\,Q_\phi Q_\varphi
$$
(see [13]). Both the expressions for $F_{\alpha}$ and $-\Delta_{\alpha}$ contain the
arbitrary parameter $\lambda$, but contrarily to the appearance, they
do not depend on it.
Indeed (following [14, \S 2]) the operator and form domain can be defined in
the following alternative, and more useful, way.
Note that, since for any $\lambda>0$
$$
G_\lambda\in \LD\,,\quad G-G_\lambda\in \HDP\,,
\quad
\left(G-G_\lambda\right)(0)=\frac{\sqrt\lambda}{4\pi}\,,
$$
where
$$G(x)={1\over {4\pi|x|}}\ ,$$
defining
$$
\phireg:=\phi_\lambda +Q_\phi(G_\lambda-G)\in\HDP
$$
we have equivalently 
\begin{eqnarray}
&&D^2_\alpha(\RE^3) =\nonumber\\
&&\left\{\phi \in\LD\, :\, \phi =\phireg+Q_{\phi}G,\
\phireg\in\HDP,\  Q_{\phi}\in\RE,\ \alpha\, Q_\phi=\phireg(0)\right\}.
\nonumber
\end{eqnarray}
Correspondingly, the form domain is
$$
D^1(\RE^3)=\left\{ \phi\in L^2(\RE^3)\ :\ \phi =\phireg+Q_\phi G,\
\phireg\in\HUP,\ Q_\phi\in\RE\right\}
$$
so that, with this definition, the singular part of the field is
exactly Coulombian. However such a singular field G {\it is not
in the configuration space} $D^1(\RE^3)$. 
The removal of this incongruence will lead, in the
following section, to the introduction 
of the operator $\bar W_\alpha$.
\par
With the domains so given we can redefine the operator and the
form as
$$
-\Delta_{\alpha}\phi= -\bar\Delta\phireg
$$
and
$$
F_{\alpha}(\phi, \varphi)=
\langle(-\bar\Delta)^{1/2}\phireg,(-\bar\Delta)^{1/2}\varphireg\rangle+
\alpha\,Q_\phi Q_\varphi\ .
$$
Now it is well known (see [11, Chap. I.1]) that $-\Delta_{\alpha}$
is a selfadjoint operator in $L^2(\RE^3)$. An important property is that $-\Delta_{\alpha}$ is positive
only for $\alpha \ge 0$, whereas for $\alpha < 0$ it is only bounded
from below; more precisely if $\alpha \ge 0$ ({\it repulsive} delta
interactions) the spectrum of the operator is absolutely continuous
and coinciding with $[0,+\infty)$; if $\alpha < 0$ ({\it attractive}
delta interactions) the spectrum is given by $\{-\lambda_0\} \cup
[0,+\infty)$, where $-\lambda_0=-(4\pi \alpha)^2$ is an isolated
negative eigenvalue, and
the remaining part of the spectrum is
absolutely continuous. In the Schr\"odinger case this eigenvalue
corresponds to a bound state, while in the wave case where one has a second
order equation in time, it leads to unstable solutions exponentially
running away in the past or in the future (see [5], [9], [10] and reference
therein for the meaning of these well known runaway solutions in classical
electrodynamics).\par
We now come to the wave generator associated to the standard delta
operator.
Its domain and action are given by
$$
W_{\alpha}:D^2_{\alpha}(\RE^3)\oplus D^1(\RE^3)\rightarrow D^1(\RE^3)\oplus
L^2(\RE^3)\ ,
$$
\begin{equation}
W_{\alpha}
\left(
\begin{array}{c}
\phi
\\
\dot \phi
\end{array}
\right)=
\left(
\begin{array}{cc}
0 & \uno
\\
\Delta_{\alpha} & 0
\end{array}
\right)
\left(
\begin{array}{c}
\phi
\\
\dot \phi
\end{array}
\right)\ .
\end{equation}
By considering the Hilbert space structure given by
$D^1(\RE^3)\oplus L^2(\RE^3)\simeq\HU\oplus\RE\oplus L^2(\RE^3)$ this operator is the generator of
a strongly continuous group of operators
$$U^t_{\alpha}:D^1(\RE^3)\oplus
L^2(\RE^3)\rightarrow D^1(\RE^3)\oplus L^2(\RE^3)\ .$$
In the case $\alpha\ge 0$
this is an immediate conseguence of the skew-adjointness of
$W_\alpha$ with respect to the positive energy scalar product on the
phase space given by
\begin{equation}
\langle\langle(\phi,\dot \phi),(\varphi,\dot \varphi)\rangle\rangle_\alpha:=
\langle\dot\phi,\dot \varphi\rangle+F_\alpha(\phi,\varphi)\ .
\end{equation}
More precisely one has the following result 
(the proof being a straighforward calculation):\vv
{\bf Theorem 2.1.} {\it For
any $\alpha\in\RE$, with respect to the scalar product 
$\langle\langle\cdot,\cdot\rangle\rangle_\beta$, $\beta\ge 0$, one has
$$D(W_\alpha^*)=D(W_\beta)$$ and
$$
W_\alpha^*(\phi,\dot\phi)=
-\left(\dphireg+\frac{\alpha}{\beta}\,Q_{\dot\phi}G,
\bar\Delta\phireg\right)\,,\qquad \beta>0\ ,
$$
$$
W_\alpha^*(\phi,\dot\phi)=
-\left(\dphireg+Q_{\dot\phi}G,
\bar\Delta\phireg\right)\equiv W_\beta(\phi,\dot\phi)\,,\qquad \beta=0=\alpha\ .
$$}
\vv
In the case $\alpha<0$ the operator $W_\alpha$ is readily proven to be a
generator by considering the operator
$$
W_{\alpha,\lambda}(\phi,\dot \phi):=(\dot \phi,(\Delta_{\alpha}
-{\lambda})\phireg)\, ,
$$
where $\lambda>{\lambda}_0$. This, being now $-\Delta_{\alpha}
+{\lambda}$ positive, is skew-adjoint with respect to the scalar
product
\begin{equation}
\langle\langle(\phi,\dot \phi),(\varphi,\dot
\varphi)\rangle\rangle_{\alpha,\lambda}
:=\langle\dot\phi,\dot \varphi\rangle+F_\alpha(\phi,\varphi)
+\lambda\langle\phi,\varphi\rangle\ ,
\end{equation}
and so it
generates a group of isometries (w.r.t. the Hilbert structure given by
(8)). The original operator $W_{\alpha}$, being a perturbation of the
previous one by a bounded operator, also generates a strongly continuous
group of operators on the phase space (which however are no more
isometries). \par 
We now describe an alternative way to prove that
$W_\alpha$, $\alpha<0$, is a generator. Such a different method will play a key role in the next sections. 
As we already said before in the case $\alpha<0$ the self-adjoint
operator $-\Delta_\alpha$ has a
negative eigenvalue $-\lambda_0$ (with corresponding normalized eigenvector
$4\pi\sqrt{-2\alpha}\,G_{\lambda_0}$) which gives rise to
the runaway solutions of the wave equation associated to
$W_\alpha$. Proceeding as in [10, \S 4]
(note that there we worked with the different decomposition
$\phi=\phi_{\lambda_0}+Q_\phi G_{\lambda_0}$) we
consider the linear operator
$$
\WNR:\DDNR\oplus\DUNR\to\DUNR\oplus\LDNR\,,
$$
$$
W_{\alpha}
\left(
\begin{array}{c}
\phi
\\
\dot \phi
\end{array}
\right)=
\left(
\begin{array}{cc}
0 & \uno
\\
\Deltanr & 0
\end{array}
\right)
\left(
\begin{array}{c}
\phi
\\
\dot \phi
\end{array}
\right)\,,
$$
where, given any vector subspace ${\cal
V}\subseteq\LD$, we have defined the corresponding ``non runaway'' subspace
$[{\cal V}]_{\nr}$ by
$$
[{\cal V}]_{\nr}:=
\{\ \phi\in {\cal V}\ :\ \langle \phi,G_{\lambda_0}\rangle=0\ \}\,,
$$
and $$\Deltanr:=(\Delta_\alpha)_{\left|\DDNR\right.}
\equiv P_\nr\cdot(\Delta_\alpha)_{\left|\DDNR\right.}\,,$$ 
$P_\nr$ being the orthogonal
projector onto $\LDNR$. By simple calculations one has (see [10, \S 4])
$$
\DUNR=\left\{\phi\in D^1(\RE^3)\ :\ Q_\phi
=-4\pi\sqrt{\lambda_0}\,\langle\phireg,G_{\lambda_0}\rangle\right\}
$$
$$
\DDNR=\left\{\phi\in D^2_\alpha(\RE^3)\ :\ \phireg(0)
=\lambda_0\,\langle\phireg,G_{\lambda_0}\rangle\right\}
$$
and
\begin{eqnarray}
&&\Deltanr\phi\nonumber\\
&=&\bar\Delta\phireg
-8\pi\sqrt{\lambda_0}\,\langle\bar\Delta\phireg,G_{\lambda_0}\rangle
G_{\lambda_0}\nonumber\\
&=&\bar\Delta\phireg+
8\pi\sqrt{\lambda_0}\,\langle(-\bar\Delta+\lambda_0)\phireg,G_{\lambda_0}\rangle
G_{\lambda_0}-
8\pi\sqrt{\lambda_0}\,\lambda_0\,\langle\phireg,G_{\lambda_0}\rangle
G_{\lambda_0}
\nonumber\\
&=&\bar\Delta\phireg\,.\nonumber
\end{eqnarray}
The {\it non-negative} bilinear form associated to $-\Deltanr$ is then
$$
F_\alpha^{\nr}(\phi,\varphi)=\langle(-\bar\Delta)^{1/2}\phireg,
(-\bar\Delta)^{1/2}\varphireg\rangle-
4\pi\,\lambda_0^{3/2}\langle\phireg,G_{\lambda_0}\rangle
\langle\varphireg,G_{\lambda_0}\rangle\,
$$
and
$\WNR$ is skew-adjoint w.r.t. the scalar product
$$
\langle\langle(\phi,\dot\phi),(\varphi,\dot\varphi)\rangle\rangle^{\nr}_\alpha:=
\langle\dot\phi,\dot\varphi\rangle+F_\alpha^{\nr}(\phi,\varphi)\ .
$$
The strongly continuous one parameter group of evolution generated by
$\WNR$ preserves the {\it non-negative} energy
$$
\E^{\nr}_\alpha(\phi,\dot\phi):=
\frac{1}{2}\,\left(\langle\dot\phi,\dot\phi\rangle+F_\alpha^{\nr}(\phi,\phi)\right)
$$
with coincides with the Hamiltonian of the sistem w.r.t. the
symplectic form $\omega$ (see [10, thm. 4.2] for an alternative
Hamiltonian picture). \par
Since $\Delta_\alpha G_{\lambda_0}=\lambda_0 G_{\lambda_0}$, and
$$
D^2_\alpha(\RE^3)\simeq\DDNR\oplus\RE\,,
$$
$$
D^1(\RE^3)\simeq\DUNR\oplus\RE\,,
$$
$$
\LD\simeq\LDNR\oplus\RE\,,
$$
we can write
$$
W_\alpha=\WNR\times \Lambda_0\,,
$$
where
$$
\Lambda_0:\RE^2\to\RE^2\,,\qquad\Lambda_0(x,\dot x)
:=(\dot x,\lambda_0\,x)\,.
$$
Therefore $W_\alpha$, $\alpha<0$, is a generator and
$$
U_\alpha^t\equiv e^{tW_\alpha}=e^{tW^\nr_\alpha}\times e^{t\Lambda_0}\,.
$$
Here and below, given two linear operators $A_1:D(A_1)\to {H}_1$ and
$A_2:D(A_2)\to {H_2}$, $A_1\times A_2: D(A_1)\times D(A_2)\to {H}_1\oplus{H}_2$ denotes the the linear operator defined by 
$$A_1\times A_2\,(\phi_1,\phi_2):=(A_1\phi_1,A_2\phi_2)\,.$$
In conclusion, for any $\alpha\in\RE$, $U^t_\alpha$ is a group of canonical
transformation w.r.t. the symplectic form $\omega$, and $W_\alpha$ is
the Hamiltonian vector field corresponding to the energy
$$
{\cal E}_\alpha(\phi,\dot\phi)=\frac{1}{2}\,\left(\,\|\dot
\phi\|_2^2+F_\alpha(\phi,\phi)\,\right)
\equiv{\cal E}(\phireg,\dot\phi)+\frac{\alpha}{2}\,Q_\phi^2\ .
$$
Let us remark that the flow $U^t_\alpha$ can be explicitly calculated
(see [5, thm. 3.1]).
\vskip20pt\p
{\bf{\sf III. THE OPERATOR $\bar W_{\alpha}$.}}\vv
Now we would like to mimic the construction of the energy space for
the usual wave generator and the extension of the operator itself, to
the case of delta point interactions.
To this end, let us define the linear operator
$$
\bar W_{\alpha}:\DDP\oplus D^1(\RE^3)\to \DUP\oplus \LD\,,\qquad
\bar W_{\alpha}(\phi,\dot \phi):=(\dot
\phi,\bar\Delta \phireg)\, ,
$$
where
$$
\DDP:=\left\{\phi=\phireg+Q_{\phi}G,\
\phireg\in\HDP,\ Q_\phi\in\RE,\ 
\alpha\, Q_{\phi}=\phireg(0)\right\}\,,
$$
$$
\DUP:=\left\{ \phi =\phireg+Q_\phi G,\
\phireg\in\HUP,\ Q_\phi\in\RE\right\}\ .
$$
Analogously to the free case $\DUP\oplus L^2(\RE^3)$ describes now the
space of finite energy states. Moreover the
Coulombian singularity $G$ {\it
is now in the configuration space} $\DUP$ .\par
Introducing the Hilbert space structure given by
$\DUP\oplus L^2(\RE^3)\simeq\HUP\oplus\RE\oplus L^2(\RE^3)$ we want now to
show that also in this case $\bar W_\alpha$ generates a strongly continuous
one parameter group of evolution. When $\alpha>0$,
considering, similarly to the case of $W_\alpha$, the scalar
product
$$
\langle\langle(\phi,\dot \phi),(\varphi,\dot \varphi)\rangle\rangle_\alpha:=
\langle\dot\phi,\dot \varphi\rangle+\langle(-\bar\Delta)^{1/2}
\phireg,(-\bar\Delta)^{1/2}\varphireg\rangle+\alpha\, Q_\phi Q_\varphi\ ,
$$
one can prove that $\bar W_\alpha$ is skew-adjoint and so it is a
generator. Note that when
$\alpha=0$, contrarily to situation discussed in the previous
section, $\langle\langle\cdot,\cdot\rangle\rangle_\alpha$ is no more
a scalar product, being annihilated by the zero energy eigenvector
$(G,0)$ (this fact has to be compared with the presence of a
zero energy resonance for $-\Delta_0$). In order to show
that also in the case
$\alpha\le 0$ $\bar W_\alpha$ is a generator 
one can not use the same strategy as before
consisting in a translation, since the scalar product (8) is now
ill-defined,
$\DUP$ being not a subset of $\LD$.
So the perturbation argument fails and we are forced to proceed in an
alternative way. The decomposition of $W_\alpha$, $\alpha<0$,
introduced at the end of the previous section is our starting point: we simply extend it to the case of $\bar W_\alpha$. Therefore we define, when $\alpha<0$,
$$
\DDPNR=\left\{\phi\in \DDP\ :\ \phireg(0)
=\lambda_0\,\langle\phireg,G_{\lambda_0}\rangle\right\}
$$
$$
\DUPNR=\left\{\phi\in \DUP\ :\ Q_\phi
=-4\pi\sqrt{\lambda_0}\,\langle\phireg,G_{\lambda_0}\rangle\right\}\,,
$$
and
$$
\BWNR:\DDPNR\oplus\DUNR\to\DUPNR\oplus\LDNR\,,
$$
$$
\BWNR(\phi,\dot\phi):=(\dot\phi, \bar\Deltanr\phi)\,,
$$
where
$$
\bar\Deltanr:\DDPNR\to\LDNR\,,
$$
$$ 
\bar\Deltanr\phi:=\bar\Delta\phireg\,.
$$
With such definitions $\BWNR$ results skew-adjoint with respect to 
the scalar product on $\DUPNR\oplus\LDNR$ given by
\begin{eqnarray}
\langle\langle(\phi,\dot\phi),(\varphi,\dot\varphi)\rangle\rangle^{\nr}_\alpha&:=&
\langle\dot\phi,\dot\varphi\rangle+
\langle(-\bar\Delta)^{1/2}\phireg,(-\bar\Delta)^{1/2}\varphireg\rangle
\nonumber\\
&&
-4\pi\,\lambda_0^{3/2}\langle\phireg,G_{\lambda_0}\rangle
\langle\varphireg,G_{\lambda_0}\rangle\,.\nonumber
\end{eqnarray}
Moreover, since
$$
\DDP\simeq\DDPNR\oplus\RE\,,\quad
\DUP\simeq\DUNR\oplus\RE\,,
$$
similarly to the case of $W_\alpha$ we have
$$
\bar W_\alpha=\BWNR\times \Lambda_0\,.
$$
For the case $\alpha=0$ a similar decomposition is possible by using
the projection onto the subspace orthogonal to
the eigenvector $(G,0)$. Indeed, defining 
$$
\bar W_{(0)}:\HDPZ\oplus D^1(\RE^3)\to\HUP\oplus\LD\,,\qquad
\bar W_{(0)}(\phi,\dot\phi):=(\dot\phireg,\bar \Delta\phi)\,,
$$
where $\HDPZ:=\left\{\phi\in\HDP\ :\ \phi(0)=0\right\}$, the operator
$\bar W_{(0)}$ is skew-adjoint with respect to the scalar product 
$$
\langle\langle(\phi,\dot\phi),(\varphi,\dot\varphi)\rangle\rangle_{(0)}
:=\langle\dot\varphi,\dot\varphi\rangle+
\langle(-\bar\Delta)^{1/2}\phi,(-\bar\Delta)^{1/2}\varphi\rangle
$$ 
and, since 
$\DUP\simeq \bar H^1(\RE^3)\oplus \RE$, the following decomposition holds:
$$
\bar W_0=\bar W_{(0)}\times 0\ .
$$
We can now state our result regarding the existence of dynamics:\vv
{\bf Theorem 3.1.} {\it $\bar W_\alpha$ is a closed operator
coinciding with the closure of $W_\alpha$. It generates a
strongly continuous group of evolution
$$\bar U^t_{\alpha}:\DUP \oplus
L^2(\RE^3) \rightarrow \DUP \oplus L^2(\RE^3)\,.$$
which can be defined as
$$
\bar U^t_{\alpha}(\phi,\dot \phi)=\lim_{n\uparrow
\infty}U^t_{\alpha}(\phi_n,\dot \phi)\,,
$$
where $\left\{\phi_n\right\}_1^\infty\subset D^1(\RE^3)$ is any sequence
such that $\phi_n\to \phi$ in $\DUP$.}
\vv
{\it Proof.} $\bar W_\alpha$ is a generator since it is skew-adjoint
when $\alpha>0$ and $\bar W_0=\bar W_{(0)}\times 0$, 
$\bar W_\alpha=\BWNR\times \Lambda_0$, $\alpha<0$,
where both $\bar W_{(0)}$ and $\bar W^\nr_\alpha$ are
skew-adjoint. Therefore $\bar W_\alpha$ is closed. By its definition
$\bar W_\alpha$ is equal to $W_\alpha$ on $D^2_\alpha(\RE^3)\oplus
D^1(\RE^3)$ and so it coincides with the closure of $W_\alpha$ if $D^2_\alpha(\RE^3)\oplus
D^1(\RE^3)$ is a core. This is proven as follows: \p 
analogously to the case of  $W_{\alpha}$, any $\phi\in \DDP$
admits the representation
$$
\phi=\phi_\lambda+Q_\phi G_\lambda
$$
where
$$
\phi_\lambda=\phireg-Q_\phi(G_\lambda-G)\in \HDP
$$
and
$$
\left(\alpha +
{{\sqrt{\lambda}}\over {4\pi}}\,\right)Q_{\phi}= \phi_{\lambda}(0)\ .
$$
Consider then a sequence $\phi^n_{\lambda}$ in
$H^2(\RE^3)$ and define
$$
\phi_n :=\phi^n_\lambda+Q_n G_\lambda\,\in D^2_\alpha(\RE^3)\ ,
$$
where
$$
Q_n:=\left(\alpha +
{{\sqrt{\lambda}}\over {4\pi}}\,\right)^{-1} \phi^n_{\lambda}(0)\ .
$$   
Now if $\phi^n_{\lambda}$ converges in $\HDP$ to $\phi_{\lambda}$,
we have that $Q_n$ converges to $Q_\phi$, thanks to the continuous
embedding of $\HDP$ in $C^0_b(\RE^3)$ (see e.g. [12, \S 5.6.2]).\par
Being $\bar W_\alpha$ equal to $W_\alpha$ on $D^2_\alpha(\RE^3)\oplus
D^1(\RE^3)$ the same is true for the corresponding groups of evolution. 
Since $D^2_\alpha(\RE^3)\oplus
D^1(\RE^3)$ is dense in $D^1(\RE^3)\oplus\LD$ which is continuously
embedded in 
$\bar D^1(\RE^3)\oplus\LD$, one has the equality $\bar U^t_\alpha(\phi,\dot\phi)=
U^t_\alpha(\phi,\dot\phi)$ for any $\phi\in D^1(\RE^3)$.
The proof is then concluded by the denseness of $D^1(\RE^3)$ in $\DUP$.\qed
\vv
Let us remark that, since $\bar W_\alpha$ is the closure of $W_\alpha$, 
our construction coincides, in the case $\alpha\ge 0$, with the
abstract one given in [3] (see also [4, \S 8] for a similar construction).
Moreover, since $W_\alpha= W_\alpha^\nr\times \Lambda_0$ when
$\alpha<0$, one has that $\bar W_\alpha^\nr$ is the closure of $W_\alpha^\nr$. 
\vskip 20pt\p
{\bf {\sf IV. THE SYMPLECTIC STRUCTURE.}}\vv
The standard symplectic structure recalled in the introduction,
$$
\omega\left((\phi,\dot \phi),(\varphi,\dot \varphi)\right):=
\langle\phi,\dot \varphi\rangle-\langle\varphi,\dot \phi\rangle
$$
it is not well defined on the phase space finite energy states, i.e
$\DUP \oplus L^2(\RE^3)$. This requires a different approach to the
Hamiltonian description of the dynamical system described in the
previous paragraph.  The problem shows up already in the case of the
free wave equation, with the phase space $\HUP\oplus L^2(\RE^3)$;
usually in the standard
literature on infinite dimensional Hamiltonian systems (see e.g. [15])
only the easier case of the free field with strictly positive mass is
explicitely discussed.\par
We recall that (see [16], [15]) when the Hilbert
space carries a complex structure $\J$, it is possible to complexify the space in
such a way that the imaginary part of the complex scalar product turns
out to be a symplectic form, while the real part is the old (real)
scalar product, coinciding with the energy. Any skew-adjoint
operator $A$ {\it commuting} with $\J$ remains skew-adjoint within the
complex Hilbert space,
so that $iA:=\J\cdot A$ is self-adjoint. Therefore, since $e^{tA}\equiv e^{-it(iA)}$,
$A$ generates a strongly continuous group of unitary (hence symplectic)
transformations. More precisely, collecting the known results on the
subject (see e.g. [15,
\S 2.6, \S 2.7], [16, chap. II]), we state the following: \vv
{\bf Theorem 4.1.} {\it Let $A$ be an injective skew-adjoint operator
on the real Hilbert space $H$ with inner product
$\langle\cdot,\cdot\rangle$.
Then the closure of the densely defined linear operator
$$A\cdot
(-A^2)^{-1/2}:\hbox{\rm Range$\,(A)$}\to{H}$$
defines a complex structure $\J$ commuting with $A$. Defining, for
any $\psi\in {H}$, the multiplication by the complex number $i$
as $$i\,\psi:=\J\psi\,,$$ $H$ becomes a complex Hilbert space with
Hermitean inner product
$$
[\psi_1,\psi_2]:=\langle\psi_1,\psi_2\rangle+i\,
\langle\psi_1,\J\psi_2\rangle\ .
$$
The strongly continuous one parameter group $U^t:=e^{tA}$ is a
group of symplectic transformations relatively to the symplectic form
$$
\Omega(\psi_1,\psi_2):=\hbox{\rm Im}\,[\psi_1,\psi_2]
$$
and the linear vector field
$$
A:D(A)\to{H}
$$
is Hamiltonian with associated densely defined Hamiltonian function
$$
\CH:D(\Q)\to\RE\,,\qquad\CH(\psi):=\frac{1}{2}\,\Q(\psi)\,,
$$
where $\Q$ denotes the quadratic form associated to the self-adjoint
operator $\J\cdot A$. }
\vv
A wide class of examples is obtained by the following construction,
which is a simple consequence of the above theorem.
Let us consider an injective
nonnegative self-adjoint operator 
$$
B:D\left(B\right)\to{K}
$$  
on the Hilbert space ${K}$ and let us consider the closure of
$$
\left(\begin{array}{cc}
\,0 &\uno\\
-B^2 &0 
\end{array}\right)
$$
on the Hilbert space ${H}=\bar
D(B)\oplus{K}$, where $\bar D(B)$ is the completion of $D(B)$
with respect to the norm $\|u\|_B:=\|Bu\|_{K}$.
In the case in which this closure is injective, the complex structure $\J$
given by the previous theorem is 
$$
\J_B:\bar D(B)\oplus{K}\to\bar D(B)\oplus{K}\,,
\qquad \J_B(u,v)
=
\left(\bar B^{-1}v,\bar Bu\right)\,,
$$
where $\bar B$ and $\bar B^{-1}$ are the closures respectively of $B$
and its inverse $
B^{-1}: \hbox{\rm Range}\,(B)\to\bar D(B)$.
This allows to endow ${H}$ with the structure of a complex Hilbert
space, which we continue to call ${H}$; 
precisely, defined a generic element as
$w:=(u,v)\in\bar D(B)\oplus{K}$,  the Hermitean scalar product
in ${H}$ is 
$$
[w_1,w_2]_B:=\langle\langle w_1,w_2\rangle\rangle_B+i\,
\langle\langle w_1,\J_B w_2\rangle\rangle_B\,,
$$
where
$$
\langle\langle w_1,w_2\rangle\rangle_B:=
\langle \bar Bu_1,\bar Bu_2\rangle+\langle v_1,v_2\rangle\,.
$$
On the product ${H}\times {H}$ we have the symplectic form
$$\Omega_B : \bar D(B)\oplus{K}\times \bar D(B)\oplus{K}\to \RE \qquad
\Omega_B \left( w_1,w_2\right)=\langle\langle w_1,\J_B w_2\rangle\rangle\,.
$$
With respect to the complex variable $w$ the wave equation 
\begin{equation}
\ddot u=-\bar B^2u
\end{equation}
assumes the Schr\"odinger-like form
$$
-i\,\dot w=\bar Bw\,.
$$
Moreover such an equation is Hamiltonian w.r.t. the symplectic form $\Omega_B$
and the densely defined Hamiltonian function
$$
{\cal H}_B : 
D(\bar B^{3/2})\times D(\bar B^{1/2})\to\RE\,,\quad
{\cal H}_B(w)=\frac{1}{2}\,\left(\|\bar B^{1/2}v\|^2+
\|\bar B^{3/2}u\|^2\right )\,,
$$
where the operator $\bar B^s$ is defined as the closure of $B^s$.\par
The strongly continuous symplectic group of operators obtained by solving the
equation (9) preserves the energy $\E_B(u,v):=\frac{1}{2}\,[w,w]_B$.
\vv
An immediate example is given by the choice
$B=\sqrt{-\Delta}:\HU\to \LD$, corresponding to the standard wave
equation and leading to the complex structure
$$
\J:\HUP\oplus\LD\to\HUP\oplus\LD\,,$$
$$
\J(\phi,\dot\phi):=
\left((-\bar\Delta)^{-1/2}\dot\phi,-(-\bar\Delta)^{1/2}\phi\right)\,.
$$
Other concrete examples are obtained  when the operator $B^2$ is a point 
interaction, more precisely $B=\sqrt{-\Delta_{\alpha}}:D^1(\RE^3)\to \LD$ with
$\alpha>0$. In this case the corresponding complex structure is given
by
$$
\J_\alpha:\DUP\oplus\LD\to\DUP\oplus\LD\,,$$
$$
\J_\alpha(\phi,\dot\phi):=
\left((-\bar\Delta_\alpha)^{-1/2}\dot\phi,-(-\bar\Delta_\alpha)^{1/2}\phi\right)\,.
$$ 
The same procedure is not directly applicable to the cases  
$\alpha \le 0$, due to the lack of skewadjointness and injectivity for the operator
$\bar W_{\alpha}$. 
A natural way out is to project the operator on the
subspace of absolute continuity and to apply the abstract
scheme to this projection. This works well for the case $\alpha < 0$,
whereas the case $\alpha =0$ deserves a different treatment. 
Here are the details of the two constructions.
\vv
In the case $\alpha<0$ we have seen in section III that 
$\BWNR$ is skew-adjoint, w.r.t. the scalar product 
$\langle\langle \cdot,\cdot\rangle\rangle^{\nr}_\alpha$, and
one-to-one. Therefore we can apply to it 
thm. 4.1 (or better the successive example with
$B=\sqrt{-\Delta^\nr_\alpha}\,\,$) obtaining the complex structure $\J_\alpha$ commuting
with $W_\alpha$, $\alpha<0$, defined as
$$
\J_\alpha:=\J_\alpha^{\nr}\times j\,,
$$
where
$$
\J_\alpha^{\nr}:\DUPNR\oplus\LDNR\to\DUPNR\oplus\LDNR\,,$$
$$ \J_\alpha^\nr(\phi,\dot\phi)
=
\left((-\bar\Deltanr)^{-1/2}\dot\phi,-(-\bar\Deltanr)^{1/2}\phi\right) $$
and 
$$ j:\RE^2\to\RE^2\,,\qquad j(x,\dot
x):=(\dot x,-x)\,. $$ 
Here,
analogously to the case $\alpha>0$, the linear operators 
$$
(-\bar\Deltanr)^{1/2}:\DUPNR\to\LDNR\,, $$ 
and 
$$
(-\bar\Deltanr)^{-1/2}:\LDNR\to\DUPNR\,, $$ 
are defined as the closures of 
$$ 
(-\Deltanr)^{1/2}:\DUNR\subset \DUPNR\to\LDNR $$
and
$$ 
(-\Deltanr)^{-1/2}:\hbox{\rm Range $\,\left((-\Deltanr)^{1/2}\right)$}
\subset \LDNR\to\DUPNR $$
respectively.\par 
We have then the complex Hilbert space of the couples 
$$
(\psi,z):=\left((\phi,\dot\phi),(x,\dot
x)\right)\in\DUPNR\oplus\LDNR\oplus\RE^2\,, 
$$ 
with the Hermitean
scalar product 
$$
[(\psi_1,z_1),(\psi_2,z_2)]_\alpha:=[\psi_1,\psi_2]_\alpha^{nr}+[z_1,z_2]\,,
$$
where
$$
[\psi_1,\psi_2]_\alpha^{nr}:=
\langle\langle\psi_1,\psi_2\rangle\rangle^{\nr}_\alpha
+i\,
\langle\langle\psi_1,\J^{\nr}_\alpha\psi_2\rangle\rangle^{\nr}_\alpha\,, 
$$ 
and
$$
[z_1,z_2]:=(z_1,z_2)+i\,(z_1,jz_2)\,,\qquad  (z_1,z_2):=\dot x_1\dot
x_2+x_1x_2\,. $$ 
The associated symplectic form is
$$
\Omega_\alpha : \DUPNR\oplus\LDNR\oplus\RE^2\to \RE \,,
$$
$$
\Omega_\alpha \left( (\psi_1,z_1),(\psi_2,z_2)\right)=
\langle\langle\psi_1,\J^{\nr}_\alpha\psi_2\rangle\rangle^{\nr}_\alpha
+(z_1,jz_2)\ . 
$$ 
With respect to the complex variables
$(\psi,z)$ the wave equation corresponding to $W_\alpha$ takes the
Schr\"odinger-like form 
$$ 
\left\{ \begin{array}{rcl}
-i\,\dot\psi&=&(-\bar\Deltanr)^{1/2}\psi\\ 
-i\,\dot z
&=&L_0z\end{array}\right.\qquad 
L_0:=\left(
\begin{array}{cc}
-\lambda_0 & 0
\\
0 & 1
\end{array}
\right)\,, 
$$ 
and such an equation is
Hamiltonian w.r.t. the symplectic form $\Omega_\alpha$ and the
densely defined Hamiltonian function 
$$
\CH_\alpha:D(\Q_\alpha^{\nr})\oplus\RE^2\to\RE\,,\qquad
{\CH_\alpha}(\psi,z)
=\frac{1}{2}\,\Q^{\nr}_\alpha(\psi)+\frac{1}{2}\,
\left(L_0z,z\right)\,, 
$$ 
where $$\Q_\alpha^{\nr}(\phi,\dot\phi)=
\frac{1}{2}\,\left(\|(-\bar\Deltanr)^{1/4}\dot\phi\|_2^2+
\|(-\bar\Deltanr)^{3/4}\phi\|^2_2\right)
$$
is the quadratic form associated to the self-adjoint operator
$\J^{\nr}_\alpha\cdot W^{\nr}_\alpha$. \par
If $(\phi,\dot\phi)\in \DUP\oplus\LD$ has the orthogonal
decomposition $(\phi,\dot\phi)\equiv \psi+z\tilde G$, where $\tilde G$
denotes the
normalized eigenvector corresponding to $\lambda_0$, then
$$
\E_\alpha(\phi,\dot\phi)=
\frac{1}{2}\,[\psi,\psi]^{nr}_\alpha+\frac{1}{2}\,(L_0z,z)\,.
$$
Therefore, being
$$\bar U_\alpha^t=e^{t\BWNR}\times e^{t\Lambda_0}$$ a strongly continuous
group of unitary and symplectic transformations, the energy is conserved by the flow.
\vv
We come now to the case $\alpha=0$.
In this case, in order to apply thm. 4.1, which requires
injectivity, it is necessary to project onto the subspace orthogonal to
the eigenvector $(G,0)$. Being $\bar W_{(0)}$ one-to-one and
skew-adjoint w.r.t. the scalar product 
$\langle\langle \cdot,\cdot\rangle\rangle_{(0)}$, 
one can then apply thm. 4.1 thus obtaining a one parameter group
$\bar U^t_{(0)}$ of symplectic transformations such that $$\bar
U^t_0=\bar U^t_{(0)}\times\uno$$
and so $\bar U^t_0$ preserves the energy 
$\E_0(\phi,\dot\phi)=\E(\phireg,\dot\phi)$.
\vv
Since $ (-\Delta_0)^{1/2}\phi =(-\bar\Delta)^{1/2}\phireg$ (note that
this equality holds true only in the case $\alpha=0$) one has
$$
\phireg=(-\bar\Delta)^{-1/2}\cdot(-\Delta_0)^{1/2}\phi
$$
and so, when $(\phi,\dot\phi)\in\HDPZ\oplus D^1(\RE^3)$,
$$
\J\cdot \bar W_{(0)}(\phi,\dot\phi)=
\left((-\bar\Delta)^{1/2} \phi,(-\Delta_0)^{1/2}\dot\phi\right)\equiv
\left((-\bar\Delta_0)^{1/2} \phi,(-\Delta_0)^{1/2}\dot\phi\right)\,.
$$
Moreover $\J$ commutes with $\bar W_{(0)}$ (see (15) in the next
section) and so $\J$ coincides with the complex structure associated to
$\bar W_{(0)}$ by thm. 4.1. With respect to the complex variable $\psi
=(\phi,\dot\phi)$ the wave equation corresponding to $\bar W_{(0)}$
assumes the Schr\"odinger-like form
$$-i\dot\psi=(-\bar\Delta_0)^{1/2}\psi\,.$$  
\vv
We summarize the results obtained in the following \vv
{\bf Theorem 4.2.} {\it For every $\alpha \in \RE \backslash \{0\}$ there exists 
a symplectic form  
$$\Omega_{\alpha}:\DUP\oplus\LD\times\DUP\oplus\LD\to\RE$$ 
with respect to which the vector
field $$\bar W_\alpha:\DDP\times D^1(\RE^3)\to\DUP\oplus\LD$$ is Hamiltoniam. 
Moreover for  $\alpha \le 0$ the analogous result occurs for the reduced vector 
fields
$$\bar W_\alpha^\nr:\DDPNR\times \DUPNR\to\DUPNR\oplus\LDNR$$
and 
$$
\bar W_{(0)}:\HDPZ\oplus D^1(\RE^3)\to\HUP\oplus\LD\ .
$$
For every $\alpha\in\RE$ the evolution group $\bar U_\alpha^t$ preserves the energy
$$\E_\alpha(\phi,\dot\phi)=\frac{1}{2}
\left(\|\dot\phi\|^2_2+F_\alpha(\phi,\phi)\right)\,.$$
}
\vskip 20pt\p
{\bf {\sf V. SCATTERING THEORY.}}\vv 
The Hilbert spaces where the operators $\bar W_\alpha$ and $\bar W$ act on,
respectively $\DUP\oplus L^2(\RE^3)$ and $\HUP\oplus L^2(\RE^3)$,
are different (also as sets), and so one is forced to use a two Hilbert
space formulation to treat scattering theory for the pair
$(\bar W_\alpha , \bar W)$. We refer to the seminal paper by Kato [4] for the relevant
constructions and results in scattering theory with two Hilbert
spaces. Our
approach will follow the lines of the construction given in [4, \S 8-9]
(also see [17, \S 3.5]).\par
From now on, given the real vector space $\LD$, we will 
denote by
${L_\C^2(\RE^3)}$ the complex vector space
$$
{L_\C^2(\RE^3)}:=
\left\{\phi_1+i\phi_2,\ \phi_1,\phi_2\in\LD \right\}\,.
$$
We begin introducing the isometries
$$
C:\HUP\oplus\LD\to L^2_\C(\RE^3)\,,
$$
$$ 
C(\phi,\dot\phi)\equiv C_0(\phi,\dot\phi):=(-\bar\Delta)^{1/2}\phi-i\dot\phi\,,
$$
$$
C_\alpha:\DUP\oplus\LD\to L^2_\C(\RE^3)\,,\quad\alpha>0\,,
$$
$$ 
C_\alpha(\phi,\dot\phi):=(-\bar\Delta_\alpha)^{1/2}\phi-i\dot\phi\,,
$$
$$
C_\alpha:\DUPNR\oplus\LDNR\to [L^2_\C(\RE^3)]_\nr\,,\quad\alpha<0\,,
$$
$$ 
C_\alpha((\phi,\dot\phi)):=
(-\bar\Delta^\nr_\alpha)^{1/2}\phi-i\dot\phi\,.
$$
These isometries lead to the following relations:
\begin{equation}
\J=C^{-1}\cdot i\,C\,,
\end{equation}
\begin{equation}
\J_\alpha=C_\alpha^{-1}\cdot
i\,C_\alpha\,,\quad \alpha>0\,,\qquad\J^\nr_\alpha=C_\alpha^{-1}\cdot
i\,C_\alpha\,,\quad \alpha<0\,,
\end{equation}
$$
\bar U^t=C^{-1}\cdot
e^{it\sqrt{-\Delta}}\cdot C\,,$$
\begin{equation}
\bar U_0^t=C^{-1}\cdot e^{it\sqrt{-\Delta_0}}\cdot C\times\uno\,,
\end{equation}
\begin{equation}
\bar U^t_\alpha=C_\alpha^{-1}\cdot e^{it\sqrt{-\Delta_\alpha}}\cdot C_\alpha\,,
\quad\alpha>0\,,
\end{equation}
\begin{equation}
\bar U^t_\alpha=C_\alpha^{-1}\cdot e^{it\sqrt{-\Delta^\nr_\alpha}}\cdot
C_\alpha\times e^{t\Lambda_0}\,,\quad\alpha<0\,.
\end{equation}
Note that the two equalities
\begin{equation}
\J=C^{-1}\cdot i\,C\,,\qquad \bar U_{(0)}^t=C^{-1}\cdot
e^{it\sqrt{-\Delta_0}}\cdot C
\end{equation}
imply, as we stated in the previous section, that $\J$ commutes with
$\bar W_{(0)}$.\par Moreover the relations (12)-(14) provide an
alternative construction of the dynamics generated by $\bar
W_{\alpha}$. In fact one could use such relations as definitions of
$\bar U^t_{\alpha}$ and then check by differentiating with respect to
the time parameter that this evolution group is generated by the 
operator $\bar W_{\alpha}$. \par We
introduce now the identification operators
$$
J_\alpha:\DUP\oplus\LD\to\HUP\oplus\LD\,,
$$
$$
J_\alpha(\phi,\dot\phi):=\left\{\begin{array}{ll} 
(\,(-\bar\Delta)^{-1/2}\cdot(-\bar\Delta_\alpha)^{1/2}\phi,\dot\phi\,)\,,
&\hbox{for $\alpha>0$}\\
(\phireg,\dot\phi)
\equiv(\,(-\bar\Delta)^{-1/2}\cdot(-\bar\Delta_0)^{1/2}\phi,\dot\phi\,)\,,
&\hbox{for $\alpha=0$}\\
(\,(-\bar\Delta)^{-1/2}\cdot(-\bar\Delta^\nr_\alpha)^{1/2}\cdot\Pi_\nr\phi,\dot\phi\,)
\,,&\hbox{for $\alpha<0$}\,,
\end{array}\right.
$$
where $\Pi_\nr$ denotes the projection
$$
\Pi_\nr:\DUP\to\DUPNR\,,
$$
and
$$
J'_\alpha:\HUP\oplus\LD\to\DUP\oplus\LD\,,
$$
$$
J'_\alpha(\phi,\dot\phi):=\left\{\begin{array}{ll}
(\,(-\bar\Delta_\alpha)^{-1/2}\cdot(-\bar\Delta)^{1/2}\phi,\dot\phi\,)\,,
&\hbox{for $\alpha>0$}\\
(\phi,\dot\phi)\,,&\hbox{for $\alpha=0$}\\
(\,(-\bar\Delta_\alpha^\nr)^{-1/2}\cdot P_\nr\cdot(-\bar\Delta)^{1/2}\phi,\dot\phi\,)
\,,&\hbox{for $\alpha<0$}\,.
\end{array}\right.
$$
We can then define the M\"oller wave operators 
$$
\Omega_{\pm}(\bar W,\bar W_\alpha;J_\alpha):=
\hbox{\rm s-}\lim_{t\to\pm\infty}\bar U^{-t}\cdot J_\alpha\cdot \bar U_\alpha^t
\cdot P_{{ac}}(\bar W_\alpha)\,,
$$
$$
\Omega_{\pm}(\bar W_\alpha,\bar W;J'_\alpha):=
\hbox{\rm s-}\lim_{t\to\pm\infty}\bar U^{-t}_\alpha\cdot J'_\alpha\cdot \bar U^t\,,
$$
where 
$$
P_{{ac}}(\bar W_\alpha):\DUP\oplus\LD\to\DUP\oplus\LD\,,$$
$$ 
P_{{ac}}(\bar W_\alpha)(\phi,\dot\phi):=\left\{\begin{array}{ll}
(\phi,\dot\phi)\,,&\hbox{for $\alpha>0$}\\ 
(\phireg,\dot\phi)\,,&\hbox{for $\alpha=0$}\\ 
(\Pi_\nr\phi,P_\nr\dot\phi)\,,&\hbox{for $\alpha<0$} \,.
\end{array}\right.
$$  
\vv
Concerning the existence of such wave operators we have the following
\vv
{\bf Theorem 5.1.} {\it The M\"oller wave operators 
$$
\Omega_{\pm}(\bar W,\bar W_\alpha;J_\alpha):=
\hbox{\rm s-}\lim_{t\to\pm\infty}\bar U^{-t}\cdot J_\alpha\cdot \bar U_\alpha^t
\cdot P_{{ac}}(\bar W_\alpha)\,,
$$
$$
\Omega_{\pm}(\bar W_\alpha,\bar W;J'_\alpha):=
\hbox{\rm s-}\lim_{t\to\pm\infty}\bar U^{-t}_\alpha\cdot J'_\alpha\cdot \bar U^t
$$
exist, are complete and are mutually adjoint isometries, i.e.
$$
\hbox{\rm Range}\,\Omega_+(\bar W,\bar W_\alpha;J_\alpha)=
\hbox{\rm Range}\,\Omega_-(\bar W,\bar W_\alpha;J_\alpha)=\uno_{\HUP\oplus\LD}\,,    
$$
$$
\hbox{\rm Range}\,\Omega_+(\bar W_\alpha,\bar W;J'_\alpha)=
\hbox{\rm Range}\,\Omega_-(\bar W_\alpha,\bar W;J'_\alpha)=
\hbox{\rm Range}\,P_{{ac}}({\bar W_\alpha}) \,,
$$
$$
\Omega_{\pm}(\bar W,\bar W_\alpha;J_\alpha)^*\cdot
\Omega_{\pm}(\bar W,\bar W_\alpha;J_\alpha)=P_{{ac}}(\bar W_\alpha)\,,
$$
$$
\Omega_{\pm}(\bar W_\alpha,\bar W;J'_\alpha)^*
\cdot\Omega_{\pm}(\bar W_\alpha,\bar W;J'_\alpha)=\uno_{\HUP\oplus\LD}\,,
$$
$$
\Omega_{\pm}(\bar W,\bar W_\alpha;J_\alpha)^*=
\Omega_{\pm}(\bar W_\alpha,\bar W;J'_\alpha)\,.
$$
}\vv
{\it Proof.} With the above definitions one has
$$
\Omega_{\pm}(\bar W,\bar W_\alpha;J_\alpha)=
C^{-1}\cdot \Omega_{\pm}(\sqrt {-\Delta},\sqrt{H_\alpha};I_\alpha) 
\cdot C_\alpha\cdot P_{{ac}}(\bar W_\alpha)\,,
$$
$$
\Omega_{\pm}(\bar W_\alpha,\bar W;J'_\alpha)=
C_\alpha^{-1}\cdot\Omega_{\pm}(\sqrt
{H_\alpha},\sqrt{-\Delta};I'_\alpha)\cdot C
$$
where
$$\Omega_{\pm}(\sqrt {-\Delta},\sqrt{H_\alpha};I_\alpha):=
\hbox{\rm s-}\lim_{t\to\pm\infty}e^{-it\sqrt {-\Delta}}\cdot I_\alpha\cdot 
e^{it\sqrt {H_\alpha}}\,,
$$
$$\Omega_{\pm}(\sqrt {H_\alpha},\sqrt{-\Delta};I'_\alpha):=
\hbox{\rm s-}\lim_{t\to\pm\infty}e^{-it\sqrt {H_\alpha}}\cdot
I'_\alpha\cdot  
e^{it\sqrt {-\Delta}}\,.
$$
Here $I'_\alpha:=P_{{ac}}(-\Delta_\alpha)$, $I_\alpha$ is its
left inverse, and
$$
H_\alpha:=\left\{\begin{array}{ll}
 -\Delta_\alpha\,,&\hbox{for $\alpha\ge 0$}\\
-\Delta_\alpha^\nr\,,&\hbox{for $\alpha<0$}\,
\end{array}\right.
$$
By Birman invariance principle one has
$$
\Omega_{\pm}(\sqrt
{-\Delta},\sqrt{H_\alpha};I_\alpha)=
\Omega_{\pm}({-\Delta},{H_\alpha};I_\alpha)
$$ and
$$
\Omega_{\pm}(\sqrt {H_\alpha},\sqrt{-\Delta};I'_\alpha)=
\Omega_{\pm}( {H_\alpha}, {-\Delta};I'_\alpha)
\,.
$$  
Therefore one has the identities
\begin{eqnarray}
\Omega_{\pm}(\sqrt
{-\Delta},\sqrt{H_\alpha};I_\alpha)&=&
\hbox{\rm s-}\lim_{t\to\pm\infty}e^{it{\Delta}}\cdot I_\alpha\cdot 
e^{it{H_\alpha}}\nonumber\\
&=&\hbox{\rm s-}\lim_{t\to\pm\infty}e^{it{\Delta}} \cdot 
e^{-it{\Delta_\alpha}}\cdot P_{{ac}}({-\Delta_\alpha})\nonumber\\
&=&\Omega_\pm(-\Delta,-\Delta_\alpha)\nonumber
\end{eqnarray}
and
\begin{eqnarray}
\Omega_{\pm}(\sqrt {H_\alpha},\sqrt{-\Delta};I'_\alpha)&=&
\hbox{\rm s-}\lim_{t\to\pm\infty} e^{-it {H_\alpha}}\cdot
I'_\alpha\cdot  
e^{-it{\Delta}}\nonumber\\
&=&\hbox{\rm s-}\lim_{t\to\pm\infty}
P_{{ac}}({-\Delta_\alpha})\cdot e^{it {\Delta_\alpha}}\cdot 
e^{-it {\Delta}}\nonumber\\
&=&P_{{ac}}({-\Delta_\alpha})\cdot
\Omega_\pm(-\Delta_\alpha,-\Delta)\nonumber\\
&=&\Omega_\pm(-\Delta_\alpha,-\Delta)\,.\nonumber
\end{eqnarray}
In conclusion one obtains the equalities
$$
\Omega_{\pm}(\bar W,\bar W_\alpha;J_\alpha)=
C^{-1}\cdot \Omega_{\pm}( {-\Delta},{-\Delta_\alpha}) 
\cdot C_\alpha\cdot P_{{ac}}(\bar W_\alpha)\,,
$$
$$
\Omega_{\pm}(\bar W_\alpha,\bar W;J'_\alpha)=
C_\alpha^{-1}\cdot\Omega_{\pm}({-\Delta_\alpha},{-\Delta};)\cdot C
$$
and the proof is concluded since the wave operators 
$\Omega_{\pm}( {-\Delta},{-\Delta_\alpha})$,
and $\Omega_{\pm}({-\Delta_\alpha},{-\Delta})$ exist, are
complete 
and are mutually adjoint isometries. This is proven (see [11, appendix E])
by the Birman-Kuroda theorem being the resolvent difference 
$$(-\Delta_\alpha+z)^{-1}-(-\Delta+z)^{-1}$$ a rank one (hence trace
class) operator.  
\qed\vv
The
previous theorem holds true also with the different
($\alpha$-independent and much simpler and natural) couple of identification
operators defined by 
$$
J:\DUP\oplus\LD\to\HUP\oplus\LD\,,\qquad J(\phi,\dot\phi):=(\phireg,\dot\phi)\,,
$$
$$
J':\HUP\oplus\LD\to\DUP\oplus\LD\,,\qquad J'(\phi,\dot\phi):=(\phi,\dot\phi)\,.
$$
This is true by [4, thms. 10.3 and 10.5] since the 
condition 10.1 in [4] is verified with $m=M=1$. In our situation such
a condition simply reads
$$
\forall\phi\in\HU\cap\hbox{\rm Range}\,P_{{ac}}({-\Delta_\alpha}) 
\,,\qquad\|\sqrt{H_\alpha}\,\phi\|_{L^2}
=\|\sqrt{-\Delta}\,\phi\|_{L^2}\,.
$$ 
In more detail one has the following\vv
{\bf Theorem 5.2.} {\it $J$ is $(\bar U_\alpha^t,\pm)$-equivalent to
$J_\alpha$, i.e.
$$
\hbox{\rm s-}\lim_{t\to\pm\infty}(J_\alpha-J)\cdot \bar
U^t_\alpha\cdot P_{{ac}}(\bar W_\alpha)=0\,,
$$
Therefore
$$
\Omega_{\pm}(\bar W,\bar W_\alpha;J):=
\hbox{\rm s-}\lim_{t\to\pm\infty}\bar U^{-t}\cdot J\cdot \bar U_\alpha^t
\cdot P_{{ac}}(\bar W_\alpha)$$
exist and are equal to $\Omega_{\pm}(\bar W,\bar
W_\alpha;J_\alpha)$. \par
$J'$ is a $(\bar U_\alpha^t,\pm)$-asymptotic left-inverse to $J$, i.e.
$$
\hbox{\rm s-}\lim_{t\to\pm\infty}(J'\cdot J-\uno)\cdot \bar
U^t_\alpha\cdot P_{{ac}}(\bar W_\alpha)=0\,,
$$
thus $$
\Omega_{\pm}(\bar W_\alpha,\bar W;J')
:=\hbox{\rm s-}\lim_{t\to\pm\infty}\bar U_\alpha^{-t}\cdot J'\cdot \bar U^t
$$ 
exist and are equal to 
$\Omega_{\pm}(\bar W,\bar W_\alpha;J)^*\equiv
\Omega_{\pm}(\bar W_\alpha,\bar W;J_\alpha')$.
}


\vskip 40pt \centerline{\sf Acknowledgments}
\vskip 5pt\p
We are grateful to Gianfausto Dell'Antonio for several
discussions and remarks.
\vfill\eject
\centerline{\sf REFERENCES}
\vskip 10pt\p
\begin{enumerate}

\item V. Petkov: {\it Scattering Theory for Hyperbolic Operators.}
Amsterdam: North Holland 1989

\item M. Reed, B. Simon: {\it Methods of Modern Mathematical
Physics. Vol. III: Scattering Theory}. New York, San Francisco,
London: Academic Press 1979

\item B. Weiss: Abstract Vibrating Systems. {\it J. Math. and Mech.}
{\bf 17} (1967), 241-255

\item T. Kato:  Scattering Theory with Two Hilbert Spaces.
{\it J. Func. Anal.} {\bf 1} (1967), 342-369

\item D. Noja, A. Posilicano: The Wave Equation with One Point
Interaction and the (Linearized) Classical Electrodynamics of a Point
Particle. {\it Ann. Inst. Henri Poincar\'e} {\bf 68} (1998), 351-377

\item G. Wentzel: {\it Quantum Theory of Fields.} New-York:
Interscience Pubbl. 1949

\item S. Coleman, R. Norton: Runaway Modes in Model Field
Theories. {\it Phys. Rev.}, {\bf 125} (1962), 1422-1428

\item E.M. Henley, W. Thirring: {\it Elementary Quantum Field
Theory}. New-York: Mc Graw-Hill 1962

\item D. Noja, A. Posilicano: On the Point Limit of the
Pauli-Fierz Model. {\it Ann. Inst. Henri Poincar\'e} {\bf 71} (1999), 425-457

\item D. Noja, A. Posilicano: Delta Interactions and
Electrodynamics of Point Particles. Published in: 
{\it Stochastic Processes, Physics and Geometry: New Interplays. II: A
Volume in Honor of Sergio Albeverio.} Providence,
Rhode Island: AMS 2000
 
\item S. Albeverio, F. Gesztesy, R. H\o egh-Krohn, H. Holden:
{\it Solvable Models
in Quantum Mechanics}. New York: Springer-Verlag 1988

\item V.G. Maz'ja: {\it Sobolev Spaces.} Berlin, Heidelberg:
Springer-Verlag 1985

\item A. Teta: Quadratic Forms for Singular Perturbations of the Laplacian.
{\it Publ. RIMS Kyoto Univ.} {\bf 26} (1990), 803-819

\item G.F. Dell'Antonio, R. Figari, A. Teta: Schr\"odinger
Equation with Moving Point Interactions in Three Dimensions. Published
in:
{\it Stochastic Processes, Physics and Geometry: New Interplays. I: A
Volume in Honor of Sergio Albeverio.} Providence,
Rhode Island: AMS 2000

\item P. Chernoff, J. Marsden: {\it Properties of Infinite
Dimensional Hamiltonian Systems. Lectures Notes in Mathematics} {\bf
425}. New York: Springer-Verlag 1974

\item I. E. Segal: {\it Mathematical Problems of Relativistic Physics.}
Providence, Rhode Island: AMS 1963

\item M.S. Birman: Existence Conditions for Wave
Operators. {\it Am. Math. Soc. Trans., Ser. 2}, {\bf 54} (1966), 91-117

\end{enumerate}
\end{document}